\documentstyle[11pt,twoside,jltp,epsfig]{article}

\title{Magneto-Transport Properties of the Rutheno-Cuprate RuSr$_2$GdCu$_2$O$_8$}

\author{J.E. McCrone, J.R. Cooper\address{IRC in Superconductivity,
University of Cambridge, Madingley Road, Cambridge, CB3 0HE, UK}
and J.L. Tallon\address{Industrial Research Ltd., P.O. Box 31310,
Lower Hutt, New Zealand}}

\runninghead{J.E. McCrone, J.R. Cooper and J.L. Tallon}
{Magneto-transport Properties of ${RuSr_2GdCu_2O_8}$}

\begin{document}

\begin{abstract}

${RuSr_2GdCu_2O_8}$ (Ru-1212) is a triple perovskite containing
both $CuO_2$ and $RuO_2$ layers.  It has attracted a great deal of
interest recently because it displays electronic ferromagnetism
and superconductivity coexisting on a microscopic scale, with
$T_{Curie}$ $\sim$ 135K and $T_c$ up to
50K\cite{Tallon2,Bernhard}.

We report magnetisation and magnetoresistance (MR) data that
exhibit effects due to the interaction between the ferromagnetic
ruthenium moments and the conduction electrons.  The MR is
negative at temperatures above $T_{Curie}$, and is proportional to
the square of the applied field well above this temperature. Below
$T_{Curie}$ the MR displays a positive peak at fields of around 15
kOe, but at high fields it becomes negative again, and
approximately linear.  We analyse the high temperature data in
terms of spin-disorder scattering theory and extract a value for
the exchange interaction between the carriers in the CuO$_2$
planes and the Ru spins.

\end{abstract}

\maketitle


\section{INTRODUCTION}
The original motivation for the synthesis of RuSr$_2$GdCu$_2$O$_8$
was to incorporate a metallic interlayer between the CuO$_2$
planes, increasing their coupling and hence their critical
current.
However, soon after the first successful synthesis \cite{Bau1} the
material was found to display not only superconductivity, but
coexisting ferromagnetism as well, first in the sister compound
$R_{1.4}$Ce$_{0.6}$RuSr$_2$Cu$_2$O$_{10-\delta}$ ($R$=Gd and
Eu)\cite{Felner1}, and then in Ru-1212
itself\cite{Tallon2,Bernhard}. Evidence has accumulated from
static magnetisation and muon spin rotation ($\mu$SR)
studies\cite{Bernhard}, and more recently from
Gd-eSR\cite{Fainstein} indicating that the two phases coexist on a
truly microscopic scale. In this paper we explore the interaction
between the transport carriers in the CuO$_2$ layers and the
ferromagnetic Ru moments. We analyse magnetisation and MR data for
the same sample in terms of the {\it s-d} model\cite{Kasuya}, and
derive a value for the exchange interaction energy.

\section{EXPERIMENTAL}
Phase-pure sintered pellets of RuSr$_2$GdCu$_2$O$_8$ were
synthesized via solid-state reaction of a stoichiometric mixture
of high-purity metal oxides and SrCO$_3$\cite{Tallon2}.  A final
extended anneal at 1060$^\circ$C in flowing high-purity O$_2$
produced a marked improvement in the crystallinity of the compound
and a corresponding lower residual resistivity (Fig
\ref{fig:reshall}a).

Bars of approximate dimensions 4$\times$0.8$\times$0.4mm were cut
using a diamond wheel and mounted on quartz substrates in a
six-contact configuration allowing both resistance and Hall
voltage to be measured simultaneously. 

Field-sweeps were performed at constant temperature by controlling
with a capacitance thermometer.  A small correction was made for
the drift in capacitance, and hence temperature, with time
(typically $<$150mK during one 0-11-0T field cycle).  A Cernox
thermometer was used to control the temperature sweeps for the
resistivity measurements, for which alternating current densities
of around 0.25Acm$^{-2}$ were used.

Magnetisation measurements were made on a sample of dimensions
$\sim$ 10$\times$1$\times$1mm using a commercial SQUID
magnetometer, with the field parallel to the long axis up to a
maximum of 50 kOe.

\section{RESULTS}

\subsection{Transport data}

\begin{figure}[th]
\begin{center}
  \epsfig{file=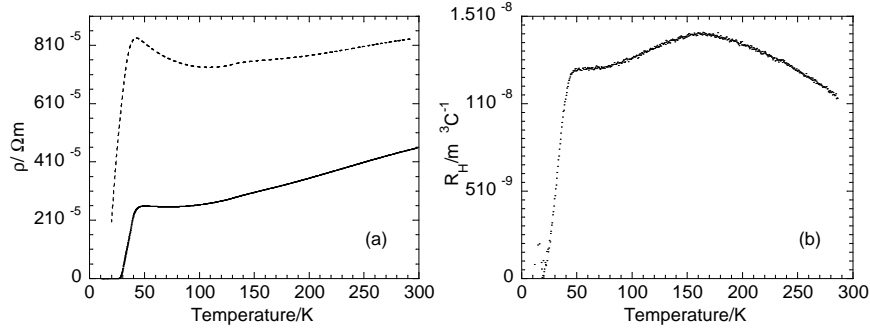, width=4.5in}
  \caption{Zero-field resistivity for as-grown (dashed) and post-anneal
  samples (a), and Hall effect (b) of sintered RuSr$_2$GdCu$_2$O$_8$.
  The Hall effect was measured in a field of 80kOe.}
  \label{fig:reshall}
\end{center}
\end{figure}

Figure \ref{fig:reshall}a shows the zero-field resistivity as a
function of temperature.  It is metallic in character at high
temperatures, with a slight upturn above $T_c$.  In this sample
both the size of the upturn, and the residual resistance have been
lowered relative to the as-grown sample by the extended anneal
mentioned above, while the TEP and Hall coefficient show little
change. This suggests that the upturn is due to grain boundary
effects.

The Hall effect data (Fig. \ref{fig:reshall}b) are similar to
those obtained for heavily under-doped polycrystalline
YBa$_2$Cu$_3$O$_{7- \delta}$, which has $R_H= 6.1 \times
10^{-9}$m$^3$C$^{-1}$ at 300K
\cite{Tony} for an oxygen deficiency, $\delta = 0.62$.  The
thermoelectric power (TEP) data are also characteristic of a
heavily under-doped high-$T_c$ cuprate, with about 0.06 to 0.07
holes per Cu atom\cite{Bernhard}.

\begin{figure}[h]
\begin{center}
  \epsfig{file=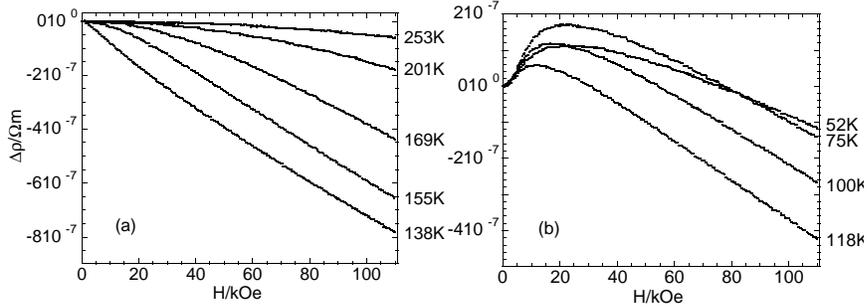, width=4.5in}
  \caption{Transverse magnetoresistance $\left( H \bot I \right)$ measured
  (a) above and (b) below $T_{Curie}$}
  \label{fig:magres}
\end{center}
\end{figure}

Figure \ref{fig:magres} shows a selection of the MR data for
temperatures above and below $T_{Curie}$.  For $T>T_{Curie}$ the
MR is always negative and it decreases as $H^2$ for temperatures
well above $T_{Curie}$.  This field dependence arises from the
freezing out of spin-disorder scattering as the Ru moments become
aligned with the magnetic field.  Close to $T_{Curie}$ the MR
becomes quite linear over the range of $H$ investigated, and for
lower temperatures displays a positive peak at 15 to 20kOe (Fig.
\ref{fig:magres}b).

\subsection{Magnetisation Data}
Magnetisation measurements were made both as a function of
temperature and field.  The temperature sweeps may be fitted using
a ferromagnetic (Curie-Weiss) plus paramagnetic (Curie) expression
of the form $\chi (T) = {C_1 \over {T- \theta}} + {C_2 \over T}$
\cite{Bernhard}.  The ferromagnetic term, $C_1$, is associated
with the Ru moments, while the Gd moments are paramagnetic.
Fitting our $\chi \left(T \right)$ data in this way for
temperatures between 150 and 300 K gives $\mu_{Ru}=1.1\mu_B$ and
$\mu_{Gd}=7.4\mu_B$, in agreement with previous
results\cite{Bernhard}. The $C_2/T$ term is later subtracted from
data obtained via the field sweeps, leaving just the ferromagnetic
component which we then compare with the MR data.

\subsection{Exchange energy}

We employ the Zener, or {\it s-d} model of a ferromagnetic metal,
in which itinerant $s$ electrons interact with $d$ electrons
localized at atomic sites. This model was extended by
Kasuya\cite{Kasuya} to calculate the MR of dilute magnetic alloys.
Assuming negligible potential scattering from the spatially
ordered Ru moments, the magnetic contribution to the resistivity
becomes\cite{Kondo}

\begin{equation}
\Delta \rho \simeq {-2 \pi N_{\epsilon_F}\over z N_V}{m \over e^2
\hbar} c J_{ex}^2 S(S+1) \alpha ^2{4 \over 27} \label{eqn:kondo}
\end{equation}

Where $N_{\epsilon_F}$ is the density of states per spin per unit
cell and $N_V$ is the number of unit cells per unit volume, each
containing $z$ conduction electrons of mass $m$. $J_{ex}$ is the
exchange interaction between the spins and conduction electrons
and $c$ is the spin concentration.

In the limit of high temperatures $\left( \alpha = {g\mu_B H\over
k_BT} \ll 1 \right)$, $\left< S_z \right> \simeq S(S+1) \alpha /3$
and the magnetisation is $M= \mu_B g \left< S_z
\right>$\cite{Beal}.  Substituting into eqn. \ref{eqn:kondo} gives
the well-known experimental result $\Delta \rho \propto M^2$ for
dilute magnetic alloys.

\begin{figure}
\begin{center}
  \epsfig{file=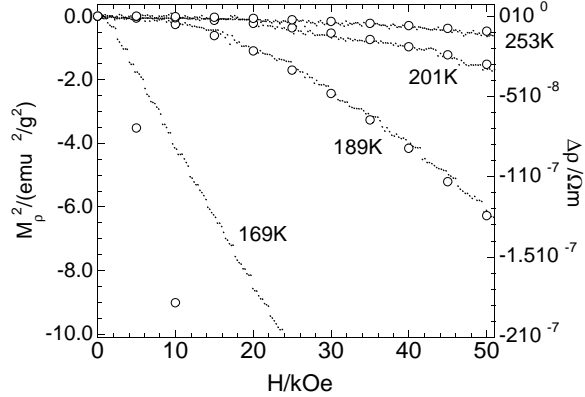, width=3in}
  \caption{Magnetisation squared (open circles) and magnetoresistance plotted
  versus applied field.}
  \label{fig:cwmagres}
\end{center}
\end{figure}

Figure \ref{fig:cwmagres} shows the magnetoresistance plotted
against magnetic field for three temperatures above T$_{Curie}$.
The square of the Ru magnetisation is also plotted, the
paramagnetic Gd component having been subtracted.

Though this material is not a dilute alloy, the {\it s-d} model is
a good fit to the data for temperatures well above T$_{Curie}$,
allowing an estimate of the exchange interaction $J_{ex}$ to be
made. The Uemura relation gives $n_s/m = 4.8\times 10^{56,}$ based
on a $T_c$ of 46K\cite{Jeff2}, and with $N_{\epsilon_F} \simeq 3
\times 10^{19}$ J$^{-1,}$ \cite{Loram} we derive an exchange
interaction of 27-47 meV depending on the choice of $c$, the
higher value being for transport solely in the CuO$_2$ planes.
These values are of the order of the superconducting energy gap
and so would be expected to have a significant effect on the
superconducting properties.

\section{CONCLUSIONS}
The above results lead us to two possible scenarios.  The first is
that the RuO$_2$ layer is a local-moment ferromagnet, the
transport occurring in the CuO$_2$ layers.  In this case it is
easy to understand the resistivity, TEP and Hall data, but
difficult to see why such a large exchange field between the
carriers and Ru moments does not seriously affect the
superconductivity.  The second scenario is that significant
current flows in the RuO$_2$ planes and the material is therefore
an itinerant ferromagnet.  In this case the MR is determined by
interactions between Ru spins and RuO$_2$ carriers, which can be
large without affecting the superconductivity.  It is, however,
more difficult to understand the resistivity, TEP and Hall data
within this scenario.

\section*{ACKNOWLEDGEMENTS}
The authors wish to thank J.W. Loram for many helpful discussions.
This work was supported by the UK EPSRC, grant RG26680.

\end{document}